\documentclass[useAMS,usenatbib]{mn2e}

\def\aap{{A\&A}}

\def\apjl{{ApJ}}

\def\apjs {{ApJS}}

\def\Msun {\,\mathrm{M}_\odot}

\newcommand{\referee}[1]{#1}

\usepackage{amsmath}
\usepackage{amssymb}
\usepackage{graphicx}
\usepackage{color}
\usepackage{url}
\usepackage{float}
\title[Detection strategies for the first SNe]{Detection strategies for the first supernovae with JWST}

\author[T. Hartwig, V. Bromm and A. Loeb]
{\parbox{\textwidth}{
Tilman Hartwig$^{1,2,3}$\thanks{E-mail: tilman.hartwig@ipmu.jp}, Volker Bromm$^4$ and Abraham Loeb$^5$\\
}\\
$^{1}$Sorbonne Universit\'es, UPMC Univ Paris 06 \& CNRS, UMR 7095, Institut d'Astrophysique de Paris, F-75014 Paris, France\\
$^2$Department of Physics, School of Science, University of Tokyo, Bunkyo, Tokyo 113-0033, Japan\\
$^3$Kavli IPMU (WPI), The University of Tokyo, Kashiwa, Chiba 277-8583, Japan\\
$^{4}$Department of Astronomy, University of Texas, Austin, Texas 78712, USA\\
$^5$Department of Astronomy, Harvard University, 60 Garden Street, Cambridge, MA 02138, USA
}

\begin{document}


\pagerange{\pageref{firstpage}--\pageref{lastpage}} \pubyear{2016}

\maketitle

\label{firstpage}

\begin{abstract}
Pair-instability supernovae (PISNe) are very luminous explosions of massive, low metallicity stars. They can potentially be observed out to high redshifts due to their high explosion energies, thus providing a probe of the Universe prior to reionization. The near-infrared camera, NIRCam, on board the {\it James Webb Space Telescope} is ideally suited for detecting their redshifted ultraviolet emission. We calculate the photometric signature of high-redshift PISNe and derive the optimal detection strategy for identifying their prompt emission and possible afterglow. We differentiate between PISNe and other sources that could have a similar photometric signature, such as active galactic nuclei or high-redshift galaxies. We demonstrate that the optimal strategy, which maximizes the visibility time of the PISN lightcurve per invested exposure time, consists of the two wide-band filters F200W and F356W with an exposure time of 600\,s. \referee{For such exposures, we expect one PISN at $z \lesssim 7.5$ per at least 50,000 different field of view, which can be accomplished with parallel observations and an extensive archival search.}
The PISN afterglow, caused by nebular emission and reverberation, is very faint and requires unfeasibly long exposure times to be uniquely identified. However, this afterglow would be visible for several hundred years, about two orders of magnitude longer than the prompt emission, rendering PISNe promising targets for future, even more powerful telescopes.

\end{abstract}


\section{Introduction}
The frontiers of astronomy are approaching the epoch when the first generation of stars and galaxies were born. These first, so-called Population~III (Pop~III), stars are the key for understanding the formation of all subsequent structures in the Universe: they lit up the Universe when its age was less than 5\% of its current age, they might have provided the seeds for supermassive black holes, and their creation of heavy elements had set the scene for the emergence of the first galaxies. Pop~III stars are believed to have formed at redshifts $z=10-25$ and to have had a higher characteristic mass than present-day stars, due to the lack of metals as efficient coolants at high redshift \citep{bromm99,abel02,yoshida03}. Their exact mass range and distribution is needed for accurately determining the ensuing radiative and chemical feedback, but it is still debated \citep{stacy10,clark11,hosokawa11,susa14,hirano14,hartwig15,stacy16,deB17,hirano17}. In addition to existing numerical simulations, it is important to obtain observational evidence to constrain the initial mass function (IMF) of the first stars.

Low-metallicity stars in the mass range $140-260\Msun$ are expected to go through a pair-instability supernova (PISN) at the end of their stellar lifetime. These very luminous explosions with energies of $10^{51-53}$\,erg are triggered by electron-positron pair production in the hot core of massive stars. The resulting reduction in radiation support causes the core to collapse, thus igniting explosive oxygen and silicon burning which completely disrupts the star, leaving no compact remnant behind \citep{rakavy67,barkat67,fraley68,bond84,fryer01}. PISNe are powered by the diffusion of the shock energy, the interaction of the ejecta with the surrounding medium, and by a major contribution from radioactive decay of nickel over a time of about one year \citep{kasen11}.
At slightly lower masses, metal-free stars can also encounter pulsational PISNe, where the pair production does not completely disrupt the star, but initiates a pulsational sequence of several core contractions, which are individually not powerful enough to unbind the star \citep{whalen13,spera17,woosley17,tolstov17}. With stellar rotation, PISNe can occur down to stellar masses of $\sim 80\Msun$, avoiding the pulsational phase \citep{chatzopoulos12,yoon12,smidt15}.
Other types of superluminous supernovae (SLSNe) were proposed in the literature as well \citep{umeda03,tominaga09,moriya10,gy12,nicholl13,yoshida14,abbott17,moriya18}. \citet{cooke12} reported the detection of two SLSNe at $z=2$ and $z=4$ with slowly evolving lightcurves. They predicted that the rate of SLSNe at higher redshift is one order of magnitude higher than in the local Universe. None of the detected SLSNe \citep{pan17,chen17,bose18} sits comfortable within the model predictions for PISNe, and alternative mechanisms for these class of SLSNe are advocated and debated \citep{dessart12,nicholl13,yusof13,lunnan17,dc17}.

The possibility of detecting PISNe at high redshift and thereby constraining the properties of their stellar progenitors has been previously discussed \citep{mackey03,scan05,whalen13b,pan13,deSouza14,wang17}. Those predictions are based on theoretical models of PISN lightcurves with multidimensional radiative transfer codes \citep{scan05,woosley07,kasen11,pan12,chen15,jerkstrand16,kozyreva17,gilmer17}. Cosmological simulations and semi-analytical models of structure formation predict the PISN rate as a function of redshift \citep{me97,wise05,hummel12,pan12,johnson13,tanaka13,magg16}, concluding that PISNe are rare events that require an optimised survey strategy and a large field of view to detect them. A PISN has not yet been observed, but a direct detection would be extremly valuable for our understanding of Pop~III star formation, because this mass range has not been probed by stellar archaeology \citep[but see][]{aoki14}.


The main focus of this paper is to find efficient detection strategies to detect PISNe at high redshift with the {\it James Webb Space Telescope (JWST)}\footnote{\url{https://www.jwst.nasa.gov}}, as well as other next-generation facilities. Based on realistic lightcurves, we determine their photometric signature and derive the optimal detection strategy by maximizing the visibility time of a PISN event for a given exposure time. We present the optimal 2-filter combination and exposure time for the prompt emission. We also discuss the possibility to detect the PISN afterglow, caused by nebular emission and reverberation in the surrounding gas. We thus introduce a novel probe of the pre-galactic medium, whose utility, however, has to await the advent of even more powerful telescopes in the future.



\section{Methodology}
The first stars are expected to form as small multiples in metal-free gas, with a distribution
that extends to high masses \citep{stacy10,clark11,greif11}. As a result, they produce a strong UV flux, which partially or completely ionizes the surrounding interstellar medium (ISM). If at least one star is in the mass range $140-260\Msun$, a PISN will be triggered. In this section, we describe how we model the ISM, the underlying stellar population, the radiative feedback on the ISM, and the interaction of the PISN with the gas. We also consider other astrophysical sources that exhibit a similar spectral energy distribution (SED), thus possibly mimicking a PISN signature.

\subsection{Single Stellar Population}
We assume that the PISN progenitor stars are embedded in a metal-free stellar cluster or group. Consequently, we also account for the radiative feedback from the underlying stellar population, resulting in the build-up of an H{\sc ii} region. To generate synthetic spectra of these primordial clusters, we consider single stellar population (SSP) models by \citet{schaerer02}, where Pop~III stars radiate close to a blackbody \citep{bromm01}. We have verified that our modeling agrees with the Pop~III SEDs by \citet{zack11}.

As a fiducial model, we use a logarithmically-flat IMF in the range $M_\mathrm{min}=3\Msun$ to $M_\mathrm{min}=300\Msun$, motivated by simulations \citep{clark11,greif11,dopcke13} and empirical hints \citep{aoki14}. Given that metal-free stars with masses of $140-260\Msun$ are predicted to explode as PISNe \citep{heger02}, we expect on average one PISN event per $500\Msun$ in Pop~III stars.
We further assume that all Pop~III stars form in one initial burst with a duration of $\Delta t_\mathrm{sb} = 10^5\,\mathrm{yr}$, which corresponds to the free-fall time at typical densities in the pre-stellar core \citep{stacy10,clark11}. This is short compared to typical stellar lifetimes of over several million years.

The rest-frame time interval over which the prompt PISN emission is bright extends for $\sim 1$\,yr, whereas the time over which stars explode as PISN, averaged over their different progenitor masses, is $\Delta t_\mathrm{PISN} \approx 10^5$\,yr. Together with $\Delta t_\mathrm{sb}$, this yields an interval of $\sim 2 \times 10^5$\,yr over which PISNe are possible for a given primordial galaxy. Consequently, one has to be lucky to observe such a galaxy at the right moment to detect a PISN signature. Even for a galaxy with a total mass of $10^5\Msun$ in Pop~III stars, and hence $\sim 200$ expected PISN explosions, the probability to observe two PISNe at the same time is negligibly small. We hence focus on the case in which one single PISN explodes at a time.

\subsection{PISN spectra}
We use the tabulated PISN spectra by \citet{kasen11}, in the form of the SED of PISNe for different times after the explosion. Specifically, we employ the model of red supergiants, since Pop~III stars in the PISN mass range are expected to have convective envelopes, which can mix the central metals with the outer hydrogen layers. The prompt emission of each PISN is brighter than the underlying stellar population for only about 1\,yr in the source frame, and previous studies have focused on the detection of this prompt emission \citep{weinmann05,frost09,hummel12,whalen13b,deSouza13,deSouza14}. Even if this time interval gets extended due to cosmological expansion by a factor of $\sim 10$, it is still very unlikely to detect this prompt emission in high-redshift surveys and we have to determine the optimal diagnostic. We consider two additional effects in our modeling, which stretch the time of possible observation to several hundreds of years in the source rest-frame: nebular emission from the ionized gas and the geometrical echo effect of the ambient medium.

\subsection{Nebular Emission}
To model the observed SED of the superposed Pop~III stellar population and a possible PISN contribution, we use the photoionization code {\sc cloudy} \citep{ferland13}, with gas of primordial composition and no dust. For the geometry and density of the ambient gas we assume three different models: a constant gas density of  $n_\mathrm{H}=1\,\mathrm{cm}^{-3}$ or $n_\mathrm{H}=100\,\mathrm{cm}^{-3}$ out to a radius of $100$\,pc, falling off as $r^{-2}$ beyond \citep{wang12}, or the idealized case of very low ambient gas density, for simplicity represented as $n_\mathrm{H}=0$. The first two models correspond to the typical conditions in a minihalo ($M_\mathrm{vir}=10^6\Msun$) and an atomic cooling halo ($T_\mathrm{vir}\gtrsim 10^4$\,K), respectively. The last model, lacking any reprocessing of the emitted radiation, corresponds to a negligibly low nebular density, caused by strong stellar feedback, or the clumpiness of the gas into a few high-density regions with a small covering fraction \citep{santos02}. The assumed density is in agreement with observations of the circumburst medium of the gamma-ray burst GRB090423 at $z=8.2$ \citep{chandra10}.
For the inner radius of the spherical gas distribution we assume 1\,pc, which corresponds to the size of the Pop~III star-forming region. We have verified that the exact choice of this value has no significant influence on the final results.

The Str\"omgren radius, within which gas is completely ionized, is given by
\begin{equation}
R_\mathrm{S} = \left( \frac{3 Q_\mathrm{H}}{4 \pi \alpha n_\mathrm{H}^2} \right)^{1/3} \approx 66\,\mathrm{pc} \left( \frac{M_*}{10^5 \Msun} \right)^{1/3} \left( \frac{n_\mathrm{H}}{100\,\mathrm{cm}^{-3}} \right)^{-2/3}
\end{equation}
with an ionizing photon flux of $Q_\mathrm{H}$, a recombination coefficient of $\alpha = 5 \times 10^{-13}\,\mathrm{cm}^3\,\mathrm{s}^{-1}$, and a total mass in Pop~III stars of $M_*$. To relate the flux of ionizing photons to the stellar mass, we use tabulated stellar evolution models \citep{schaerer02}, averaged over the IMF, whereas the exact upper limit of the IMF for this average has only a minor influence \citep{bromm01}. We note that the Str\"omgren radius is smaller than the constant density core for a gas density of $n_\mathrm{H}=100\,\mathrm{cm}^{-3}$, but larger for a gas density of $n_\mathrm{H}=1\,\mathrm{cm}^{-3}$. This indicates that the gas is already completely ionized when a PISN explodes in a minihalo, and we do not expect a significant additional contribution from nebular emission in this case \citep{zack11}.

\begin{figure}
\centering
\includegraphics[width=0.47\textwidth]{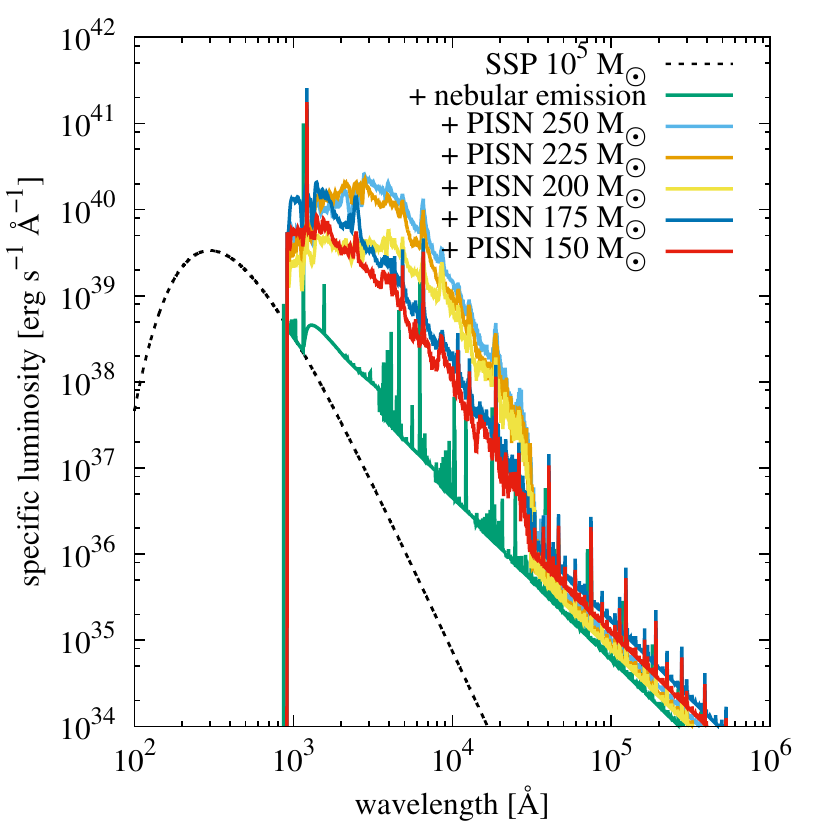}
\caption{SEDs of a SSP of $10^5\Msun$, with and without PISN contribution, and an ambient gas density of $n_\mathrm{H}=100\,\mathrm{cm}^{-3}$. The PISN SEDs are averaged over the time of explosion, weighted by their bolometric luminosity. Photons bluewards of $912$\,\AA\,are absorbed by the neutral surrounding medium, and re-emitted in the IR.}
\label{fig:cloudy}
\end{figure}

The SED of a single stellar population (SSP) with and without the contribution of different PISNe can be seen in Fig. \ref{fig:cloudy}.
The prompt PISN emission is about two orders of magnitude brighter than the underlying Pop~III stellar population.

\subsubsection{Recombination}
\begin{figure}
\centering
\includegraphics[width=0.47\textwidth]{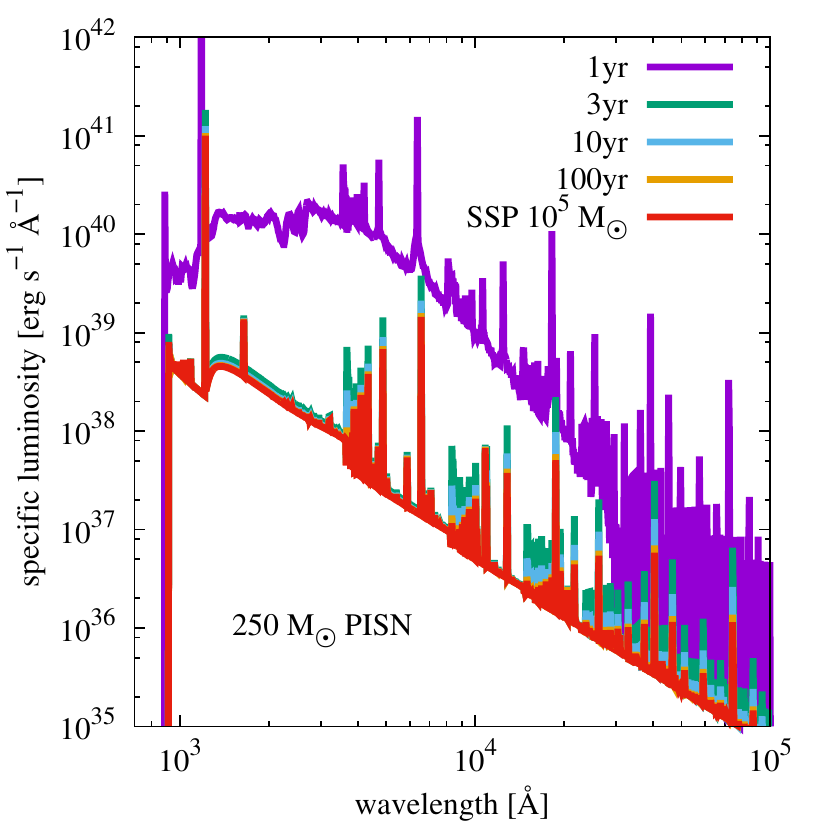}
\caption{Spectral evolution in the PISN host. SEDs of a $10^5\Msun$ SSP (red) and the time evolution of the SED after one PISN with a progenitor mass of $250\Msun$ explodes. Initially, the PISN is about two orders of magnitude brighter than the stellar cluster, but its contribution rapidly decays, over several hundred years, back to its original value.}
\label{fig:cloudyt}
\end{figure}

The nebular emission caused by the PISN decays over the recombination time scale of $t_\mathrm{rec} \sim 400$\,yr at $n_\mathrm{H}= 10^2\,\mathrm{cm}^{-3}$.
This time scale is between the characteristic time scale of the PISN light curve ($\lesssim 1$\,yr) and the typical stellar lifetimes of several Myr. We can consequently assume that the energy input from the PISN is almost instantaneous and that the SED of the underlying stellar population does not significantly change over $t_\mathrm{rec}$. We model the nebular emission with {\sc cloudy}, assuming a time-dependent incident radiation field. The resulting time evolution of the SED can be seen in Fig. \ref{fig:cloudyt}.
Already 3\,yr after the explosion, the prompt emission decayed and only a few recombination lines render the PISN afterglow brighter than the underlying stellar population. The additional nebular emission from the PISN derives almost exclusively from recombination lines, with only a small contribution of free-free emission to the continuum.

\subsubsection{Geometrical echo effect}
The light-crossing time for a spherical gas cloud with radius $R=100$\,pc is $\sim 650$\,yr, which is comparable to the recombination time. Consequently, we also have to take into account geometrical effects of the ambient gas and the reverberation that is created by photons that are absorbed and re-emitted by gas off the line of sight. We assume a central energy source that emits isotropically in all directions, embedded in a spherical ambient medium and an observer at infinity. Photons get absorbed and might be re-emitted into the direction of the observer, but they will arrive later at the observer than photons that are directly emitted into the observer's direction. The locus for which photons have the same travel time to the observer is a paraboloid.
The cross-sectional area of this paraboloid with a sphere of gas of radius $R$ as a function of time is given by
\begin{equation}
A(t) = \frac{8 \pi}{3} \sqrt{ct} \left( R-\frac{ct}{2} \right) ^{3/2}.
\end{equation}
The observed SED is a convolution of SEDs emitted at different times
\begin{equation}
F_\lambda (t) = \int _0 ^{2R/c} \mathrm{d}t' F_\lambda (t-t') N(t'),
\end{equation}
where $N(t')$ represents the surface of the paraboloid, normalized over the convolution interval.
We add this echo effect in post-processing and do not treat it self-consistently, because a proper treatment would require full 3D radiation-hydrodynamics simulations, which is beyond the scope of this paper.

\subsubsection{Chemical and Mechanical feedback}
Besides the radiative input, a PISN will also chemically enrich the ISM and affect the gas density by mechanical feedback. However, the time until the metal-rich shock front reaches the Str\"omgren radius and could affect the nebular emission is $\gtrsim 2500$\,yr for a gas density of $n_\mathrm{H}=100\,\mathrm{cm}^{-3}$ and a shock velocity of $\lesssim 0.1c$ \citep{kasen11}. Consequently, chemical and mechanical feedback do not affect the nebular emission for at least 2500\,yr after the explosion of a PISN.

\subsection{NIRCam filters}
NIRCam, the near-infrared (NIR) photometer on {\it JWST} has one short-wavelength channel and one long-wavelength channel, which together cover the range of $0.5-5\,\mu$m. This makes NIRCam the perfect instrument to probe high-redshift galaxies, whose strong rest-frame UV emission gets shifted into the NIR in the observer frame.
For the calculation of the luminosity distance, we assume a flat Universe with cosmological parameters determined by the {\it Planck satellite} \citep{planck16}.


\section{photometric signatures}
\subsection{SEDs of high-redshift sources}
We first present the resulting SEDs of the PISNe, their time evolution, and the effects of nebular emission and reverberation. In addition, we construct the SEDs of other sources at high redshift, which could have the same photometric signature as PISNe and hence confuse their detection.

\subsubsection{PISNe}
For the low density case, and that without any ambient medium, there is no additional afterglow or echo effect because the underlying SSP already keeps all the gas ionized \citep{zack11}. We only expect to see the prompt emission in these cases over a typical time scale of one year.

\begin{figure}
\centering
\includegraphics[width=0.47\textwidth]{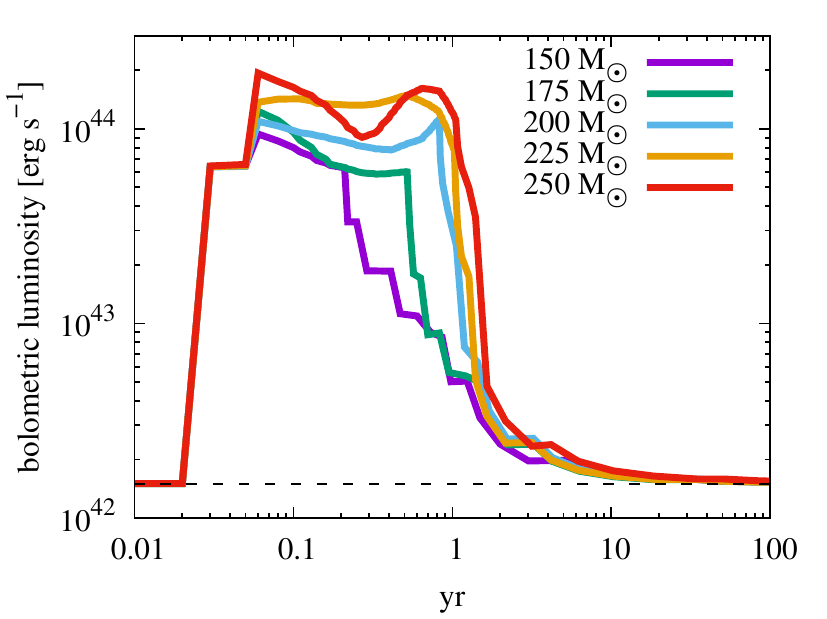}
\includegraphics[width=0.47\textwidth]{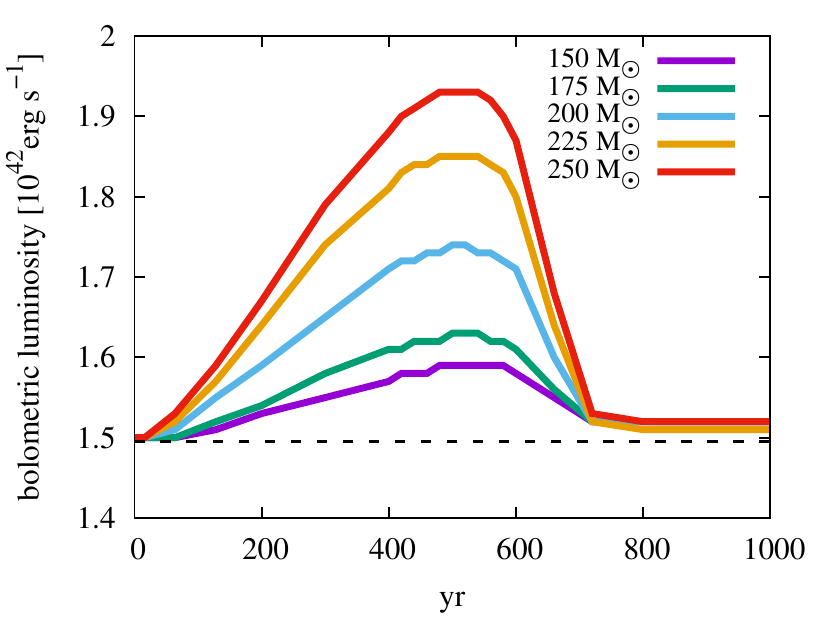}
\caption{Bolometric light curves of the total emission after the explosion of PISNe of different progenitor masses. The time is source rest-frame and measured from when the first radiation escapes into the IGM. The inclusion of nebular emission (top) stretches the light curves to $\gtrsim 10$\,yr. The additional echo effect (bottom) smooths the light curves, which has two important consequences: the bolometric luminosity is fainter, but PISNe outshine the underlying stellar population for almost one thousand years, significantly longer than without this effect.}
\label{fig:Lbol}
\end{figure}

In Fig. \ref{fig:Lbol}, we compare the bolometric light curves after SN explosion with and without the echo effect.
The reprocessing of the prompt emission stretches the lightcurve from $\sim 1$\,yr to $\gtrsim 10$\,yr and hence increases the time interval of possible detection. However, the recombination line emission decays exponentially towards the contribution from the nebular emission powered by the underlying stellar population. This yields luminosities of the reprocessed radiation that are brighter by only a factor of a few than the stellar population for $\gtrsim 1$\,yr after the PISN explosion. The echo effect smooths the lightcurve over the light crossing time of the ISM. Although this increases the time interval in which we can detect PISNe by more than two orders of magnitude, the luminosity of the PISN lightcurve is reduced by about two orders of magnitude due to conservation of the explosion energy (Section \ref{sec:glow}).

\subsubsection{AGN}
The typical PISN (averaged over progenitor masses and time after explosion) has a luminosity of $\sim 10^{43}\,\mathrm{erg}\,\mathrm{s}^{-1}$ and peaks at $\sim 200$\,nm. An AGN with a BH mass of $\sim 10^{7}\Msun$ peaks in the same wavelength range and has the same luminosity as a typical PISN for an Eddington rate of $1\%$ \citep{volonteri17}. We model the emission from a typical broad line region (BLR) illuminated by such an AGN with {\sc cloudy}, assuming a density of $n_\mathrm{H}=10^{10}\,\mathrm{cm}^{-3}$ and solar metallicity for the BLR.

Direct collapse black holes (DCBHs) are an advocated sub-type of high-redshift quasars that form due to the isothermal collapse of a pristine protogalaxy under the influence of a nearby photodissociating source \citep{bromm03b}. The isothermal collapse of pristine protogalaxies can yield seed BHs with masses of up to $\sim 10^5\Msun$, which can then further accrete gas to become more massive. These accreting DCBHs might also be detectable with {\it JWST} \citep{pacucci16,natarajan17}, but due to their small number density \citep{habouzit16}, we do not explicitly account for them, because they will not be the most abundant sources observed with {\it JWST}.

\subsubsection{Galaxies}
We also verify that our proposed photometric diagnostic is suited to distinguish a PISN from a typical high-redshift galaxy. We use the SEDs by \citet{barrow17}, who present synthetic observations of $z \geq 6$ galaxies, based on the Renaissance simulation. They demonstrate that there is no ``typical'' high-redshift galaxy, but their spectra vary with stellar mass, metallicity, gas mass fraction, formation history, and viewing angle. To account for these effects, we use averaged spectra over all galaxies with a stellar mass in the range $5.5 < \log(M_*/\Msun) < 6.5$ and assume an average bolometric luminosity of $5 \times 10^{41}\,\mathrm{erg}\,\mathrm{s}^{-1}$. This mass range is the best trade-off between the abundance and luminosity of high-redshift galaxies: low-mass galaxies are more abundant and hence more typical at high redshift. However, the faintest galaxies are below the sensitivity threshold of {\it JWST} and despite their abundance they will not be the most frequently observed sources with {\it JWST}. Galaxies with $M_*\approx 10^6\Msun$ have a comoving number density of
\begin{equation}
\frac{\mathrm{d}n}{\mathrm{d}\log M_*} \approx 10^{-3}\,\mathrm{cMpc}^{-3}
\end{equation}
at redshift $z=6$ \citep{barrow17}.
The SED of a typical, metal-enriched high-redshift galaxy is close to the SED of our Pop~III SSP with roughly the same luminosity. Due to the averaging over several galactic SEDs, their emissions lines are not as prominent as for the Pop~III SSP or the PISNe with nebular emission. Once we find a diagnostic to photometrically distinguish PISNe and the underlying Pop~III SSP, we expect metal enriched galaxies not to compromise this diagnostic. We will discuss this distinction and the corresponding constraints more quantitatively in the next sections. For a more general discussion on the {\it JWST} signatures of high-redshift galaxies, see \citet{zack17}.

\subsection{Detection strategy for prompt emission}
Many previous studies focused on the modeling and observational signatures of the prompt PISN emission \citep{scan05,weinmann05,frost09,kasen11,pan12,hummel12,whalen13b,deSouza13,deSouza14}. In this section, we present the photometric signature of PISNe and for the first time derive the optimal filter combination to detect their prompt emission and determine the most efficient exposure time and detection strategy.

The basic idea and strategy to detect PISNe at high redshift is as follows: PISNe are very bright over a period of roughly one year in the source rest frame. This time interval gets stretched to about one decade in the observer frame due to the cosmic expansion. If we could identify a PISN candidate in one observation, we can observe the same source a few months later to verify if it is a transient source, which would confirm the detection of a PISN. The crucial point is to maximize the fraction of the lightcurve that we can observe, because the probability to find a PISN in the first place is directly proportional to the visibility time, $t_\mathrm{vis}$, over which we can detect it.

\begin{figure*}
\centering
\includegraphics[width=0.98\textwidth]{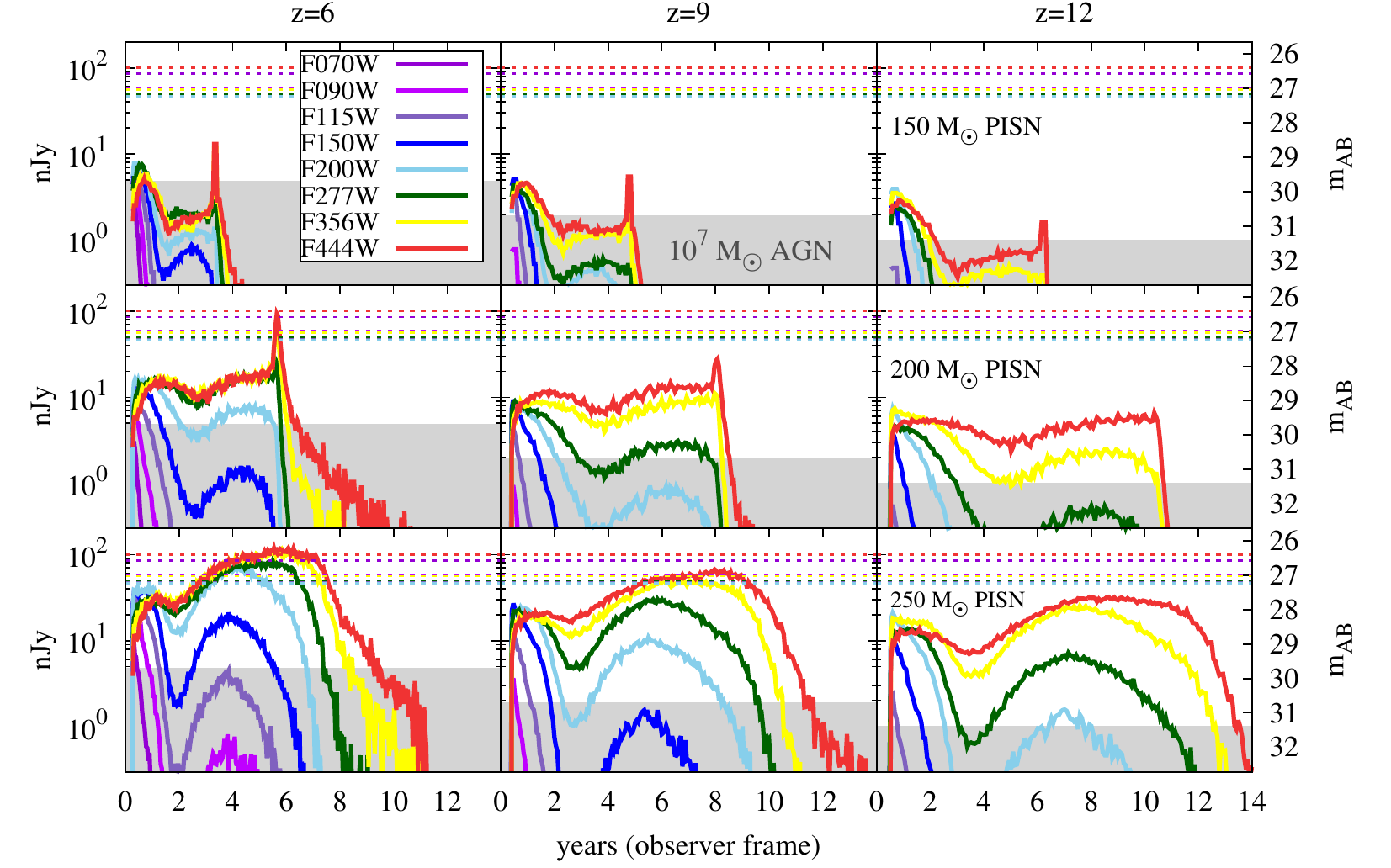}
\caption{Lightcurves of the prompt PISN emission for three different progenitor masses (top: $150\Msun$, middle: $200\Msun$, bottom: $250\Msun$) at three different redshifts (left: $z=6$, middle: $z=9$, right: $z=12$) in the {\it JWST} wide-band filters as a function of time in the observer-frame. The left vertical axis indicate the flux in nJy and the right axis yields the corresponding monochromatic AB-magnitude. The different filters are illustrated by the different colors and the corresponding sensitivity limit for a 600\,s exposure are indicated by the horizontal dotted lines in each panel. The grey area indicates the expected flux from an AGN, powered by a $10^7\Msun$ BH. For simplicity we only plot the AGN flux in the 444W filter, which is the highest flux and hence represents a conservative upper limit.}
\label{fig:trans3}
\end{figure*}

In this section, we only focus on the prompt emission and do not account for the nebular emission or echo effect. We cut the SED below $\lambda < 912$\,\AA\,to mimic the neutral IGM at high redshift, but this has only a minor influence on the calculated photometric fluxes. Due to the very faint fluxes, the wide band filters of JWST are the best choice for the photometric detection of PISNe. The expected light curves in the observer frame can be seen in Fig. \ref{fig:trans3}.
The flux increases with progenitor mass and with decreasing redshift. The time evolution of the flux is not monotonous with time and differs from filter to filter. The flux is very faint in the 070W and 090W filter, because we only have the direct prompt emission and no reprocessed recombination lines in this wavelength range. The filters with the highest expected signal-to-noise ratio (SNR) to detect PISNe are: F150W, F200W, F277W, F356W, F444W. The typical high-z galaxy and AGN yield fainter fluxes, although an AGN at $z\approx 6$ has the same flux in certain filters as a PISN at $z\approx 12$. 

We are interested in a reliable diagnostic to find PISN candidates that can then be further analysed for their transient nature. We want to maximize the observer frame time over which a PISN is detectable for a given diagnostic, because the probability to find a PISN is linearly proportional to this visibility time and since PISNe are very rare events (see Section \ref{sec:rates}), we need to identify an efficient filter combination. In the following sections, we derive the optimal detection strategy by maximizing the visibility time of the PISN lightcurve per exposure time invested. This visibility time depends on the selected filter, progenitor mass, redshift, and the desired signal-to-noise threshold. To calculate the visibility time, we define our standard sample of 5 PISNe ($150,175,200,225,250\Msun$) at 3 different redshifts ($z=6,9,12$). We then identify the observer frame visibility of this ensemble of 15 PISNe as the cumulative time over which their prompt emission is detectable in a certain filter combination. Our goal is to detect a PISN, irrespective of its progenitor mass or redshift and we hence aim at maximizing the visibility time of this entire ensemble.

To better illustrate the physical meaning of this observability time, we also provide the corresponding fraction of the lightcurve that is observable. In the optimal case, we can observe 100\% of the lightcurve, but as we will see below, this requires unnecessary long exposure times. This fraction of the lightcurve depends on our definition of the duration of the prompt emission. To simplify the comparison, we use the $250\Msun$ as standard scenario, which has a maximum visibility time of $\sim 340$\,d in the source rest frame. All following fractional visibility times of the lightcurve are for a $250\Msun$ PISN, normalized to this value.

NIRCam on board {\it JWST} has a beam splitter that allows simultaneous observation of a source in one short wavelength and one long wavelength channel. In the following sections, we make use of this feature by combining two corresponding filters.
The tabulated sensitivity $f_\mathrm{lim}$ of NIRCam is given at a SNR of S/N$=10$ and for an exposure time of $t_\mathrm{exp}=10$\,ks. These values are for point sources and the actual signal-to-noise depends on the galactic foregrounds and therefore on the source position. Our derived exposure times at S/N$=10$ should hence be treated as an optimistic lower limit. To convert the tabulated to an arbitrary exposure time, we assume the proportionality
$f_\mathrm{lim} \propto t_\mathrm{exp}^{-1/2}$
and a linear response of the detector over its dynamical range.

\subsubsection{2-filter diagnostic}
\begin{figure*}
\centering
\includegraphics[width=0.99\textwidth]{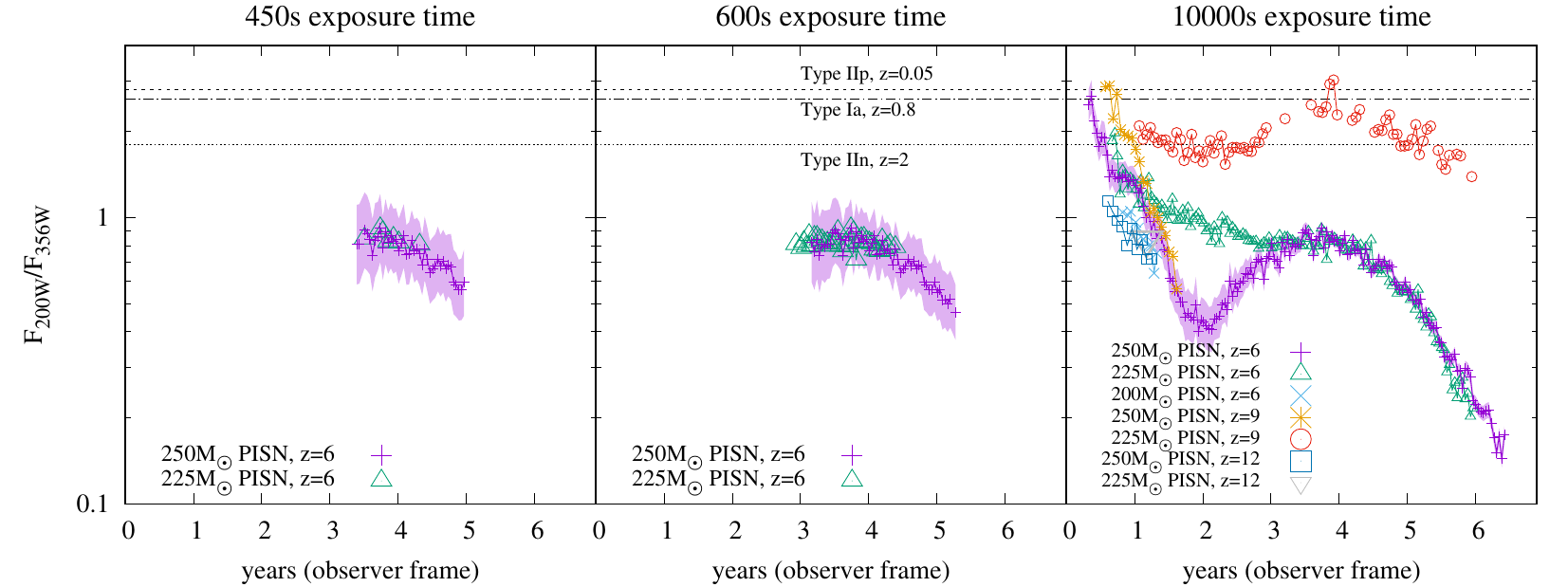}
\caption{Most promising 2-filter diagnostic for different exposure times. The points illustrate the time evolution of the light curve (connected points are separated by 2\,days in the source rest frame) for detections with signal-to-noise ratio S/N$>10$ in both filters. \referee{The shaded purple area illustrates the observational uncertainty of $<0.14$\,mag (for clarity only for the $250\Msun$ PISN at $z=6$)}. With increasing exposure time, we can see the prompt emission over a longer time period. Exposure times of $<1$\,ks are sufficient to detect the brightest PISNe at $z=6$, and for $t_\mathrm{exp}\geq 10$\,ks, PISNe at $z>12$ are detectable. \referee{To demonstrate that this photometric diagnostic is distinguishable from other SN types, we also plot the signature of three exemplary types at peak luminosity.}}
\label{fig:trans2D2f}
\end{figure*}

The most efficient diagnostic is based on two filters that can be exposed simultaneously. In Fig. \ref{fig:trans2D2f}, we show one possible 2-filter diagnostic (which we demonstrate later to be the optimal filter combination) to illustrate the effect of a varying exposure time.
As expected, the PISN visibility time increases with the exposure time. Already in a 450\,s exposure we can identify the most luminous $225\Msun$ and $250\Msun$ PISN at $z=6$ with S/N$>10$. With a longer exposure time of $\gtrsim 10$\,ks we can even detect the $200\Msun$ PISN at $z=6$ and the $225\Msun$ out to $z=12$. \referee{Also the observational uncertainty decreases with the exposure time.} In order to not only maximize the PISN visibility, but to optimise the effort, we maximize the ratio of the visibility time to the exposure time to obtain the highest probability of finding PISNe per invested exposure time. PISNe with $M<200\Msun$ are not detectable with S/N$>10$ for $t_\mathrm{exp} \leq 10$\,ks, but this does not affect the performance of the diagnostic. PISNe at higher redshift are favorable because their lightcurve gets stretched by a larger factor, which increases their observer frame visibility, but they are also significantly fainter.

\referee{We compare our photometric diagnostic to other SN types. For this comparison we select a Type~Ia SN based on the theoretical model by \citet{blondin13}, Type~IIn based on the observation of the SLSN 2008am \citep{chatzopoulos11}, and the Type~IIp SN 1999em \citep{leonard02} at their peak luminosity and illustrate them with the dash-dotted, dotted, and dashed lines in Fig. \ref{fig:trans2D2f}, respectively. The first criterion to differentiate SNe is their absolute flux in the individual filters. For our optimal diagnostic with the most promising $250\Msun$ PISN at $z=6$ we expect fluxes of $46-70$\,nJy in the F200W filter and of $55-119$\,nJy in the F356W filter. None of the other three SN types fulfills both flux criteria simultaneously at any redshift. For example, the Type IIn SN has a flux in this range in the F200W filter at a redshift of $z=2.3-3.1$ and the corresponding flux in the F356W filter at redshifts of $z=0.8-1.6$. To illustrate the flux ratio, we therefore choose the intermediate redshift of $z=2$. Our PISNe appear redder than other SNe at lower redshift. Although PISNe can be photometrically distinguished from these selected SNe at peak brightness we note that this does not cover all classes and variations of SNe and we further discuss this challenge in section \ref{sec:caveats}.}

For different filter combinations and various exposure times we provide the cumulative observer frame visibility of the PISNe in Table \ref{tab:visibility2f}.
\begin{table}
 \centering
 \caption{Observer frame visibility in days for different  2-filter diagnostics and various exposure times. The percentage in parenthesis below each visibility time indicates the fraction of the lightcurve for the $250\Msun$ PISN at $z=6$ that is visible for this exposure time.}
   \label{tab:visibility2f}
 \begin{tabular}{|c|c|c|c|c|c|c|}
 \hline
   &150W&150W&150W&200W&200W&200W\\
 $t_\mathrm{exp}$ & - & - & - & - & - & -\\
 $[\mathrm{ks}]$&277W&356W&444W&277W&356W&444W\\
  \hline 
  0.45& 0 & 0 & 0& 532 & 672 & 0\\
  &(0\%)&(0\%)&(0\%)&(22\%)&(24\%)&(0\%)\\
  0.54& 0 & 0 & 0& 700 & 1078 & 14\\
  &(0\%)&(0\%)&(0\%)&(26\%)&(28\%)&(1\%)\\
  0.6& 0 & 0 & 0& 994 & 1288 & 56\\ 
  &(0\%)&(0\%)&(0\%)&(29\%)&(32\%)&(1\%)\\
  0.72& 0 & 0 & 0& 1498 & 1526 & 322\\ 
  &(0\%)&(0\%)&(0\%)&(33\%)&(34\%)&(14\%)\\
  1.2& 0 & 0 & 0& 2268 & 2184 & 840\\ 
  &(0\%)&(0\%)&(0\%)&(44\%)&(44\%)&(34\%)\\
  3.6& 1788 & 1540 & 826 & 3746 & 3458 & 2562\\ 
  &(16\%)&(12\%)&(4\%)&(77\%)&(74\%)&(57\%)\\
  5.4& 2274 & 2070 & 1358 & 4276 & 3992 & 2940\\ 
  &(31\%)&(29\%)&(16\%)&(83\%)&(81\%)&(63\%)\\
  10& 3216 & 2848 & 2156 & 6784 & 6434 & 4786\\ 
  &(45\%)&(45\%)&(37\%)&(94\%)&(94\%)&(86\%)\\
  100& 7138 & 7188 & 5928 & 14608 & 14732 & 13326\\ 
  &(73\%)&(73\%)&(72\%)&(100\%)&(100\%)&(100\%)\\
  \hline
  \end{tabular} 
\end{table}
The visibility time for $t_\mathrm{exp} \lesssim 10$\,ks is dominated by the $225\Msun$ and $250\Msun$ PISNe at $z=6$. We assume that all PISNe have the same probability to occur and that the rate of PISNe is constant in the redshift range $z=6-9$, which agrees with \citet{magg16}; however, see \citet[][]{hummel12,johnson13}. For a different IMF or PISN rate, one could weight the contributions to the rest frame visibility accordingly. Moreover, we do not explicitly take other sources into consideration, such as metal enriched galaxies or an AGN, because they are fainter than the prompt PISN emission and vary on much longer timescales than the PISN.
The visibility fraction is an additional illustration to demonstrate the effect of a varying exposure time. It saturates for exposure times $\gtrsim 10$\,ks and doubling the exposure time does not yield much more information.

\begin{figure}
\centering
\includegraphics[width=0.47\textwidth]{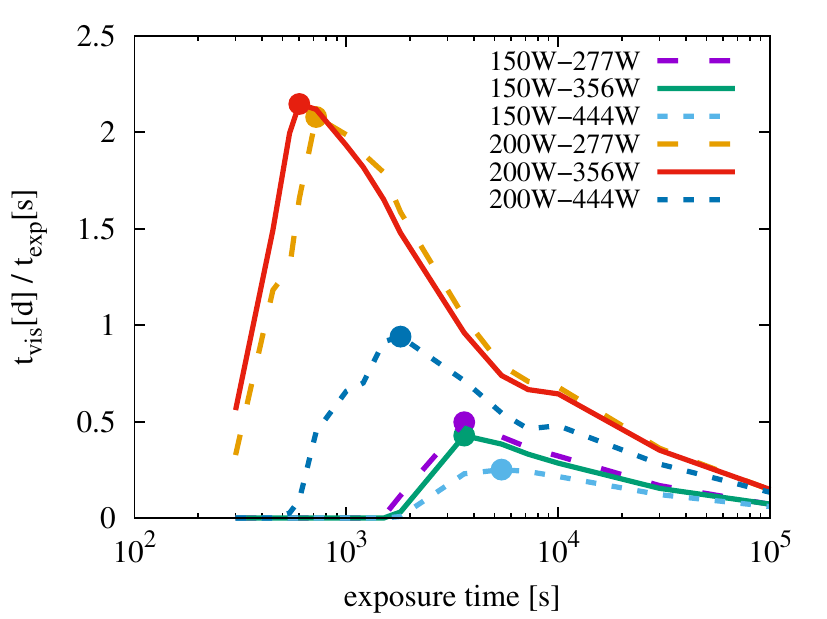}
\caption{Efficiency of 2-filter combinations, quantified by the visibility time divided by the exposure time, as a function of the exposure time. This analysis illustrates that 200W-356W is the most promising filter combination with an optimal performance for an exposure time of 600\,s. The 200W-277W color has a slightly weaker performance with almost the same optimal exposure time.}
\label{fig:texp2f}
\end{figure}

We further quantify the ratio of the visibility to the exposure time in Fig. \ref{fig:texp2f}.
We find a peak of the efficiency for the filter combination 200W-356W at an exposure time of 600\,s. With this optimal combination of filters and exposure time we can detect roughly two days of the PISN lightcurves in the observer frame per one second of exposure time. The 200W-277W color has a comparable but slightly weaker performance, and for other filter combinations the most economic exposure time is $\gtrsim 1000$\,s, but their overall performance is significantly smaller.

The filters 200W and 356W are in the short and long wavelength channel of NIRcam and can hence be used simultaneously. Our results imply that ten individual observations of different fields of view (FoV) with $600$\,s exposure time each are more promising than one with an exposure time of $6000$\,s.

\subsubsection{4-filter diagnostic}
\label{sec:4f}
The identification of PISNe as transient sources poses two challenges: first, we have to identify them as promising candidates and second, we have to verify their transient nature and their unique colors in a follow-up observation. Although the PISN SED changes with time, some colors are constant over $\gtrsim 4$\,yr (see Fig. \ref{fig:trans2D2f}), which makes it difficult to identify their transient nature with a two filter diagnostic. Moreover, we cannot account for all astrophysical sources that could have a similar photometric flux, and possibly even time evolution, as the PISN. Therefore, we also provide a color-color diagnostic based on 4 filters, which, shows a more reliable time evolution of the colors and helps to better distinguish the PISN from other sources.

\begin{figure}
\centering
\includegraphics[width=0.47\textwidth]{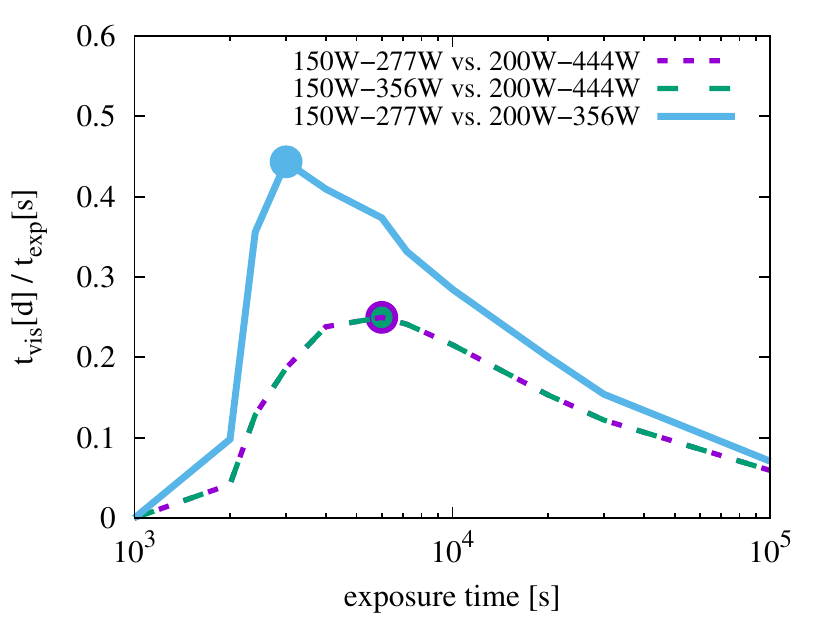}
\caption{Efficiency of 2-color diagnostics, measured by the visibility time divided by the exposure time, as a function of the exposure time. This analysis illustrates that the 2-color diagnostic 150W-277W vs. 200W-356W is the most promising filter combination with an optimal performance for an exposure time of 3\,ks. For other filter combinations including 444W the most economic exposure time is 6\,ks, but the overall performance is smaller than for the optimal filter combination.}
\label{fig:texp}
\end{figure}

Since we require that the PISN has to be detected in 4 filters simultaneously at S/N$>10$, the visibility time is shorter than in the 2-filter diagnostic. By maximizing the ratio of visibility time to exposure time we determine the optimal filter combination and corresponding exposure time (see Fig. \ref{fig:texp}).
We find a peak of the efficiency for the two color combination 150W-277W vs. 200W-356W at an exposure time of 3\,ks. With this optimal combination of filters and exposure time we can detect $\sim 11$\,h of the PISN lightcurves in the observer frame per one second of exposure time. Note that the stated exposure times are for one filter only. If the short and long wavelength channel are used at the same time, the stated exposure times have to be multiplied by two for the overall exposure time for this 4-filter diagnostic.

\subsubsection{Only $z \geq 9$ PISNe}
Our previous estimates depend on the assumptions that the probability of a PISN explosion is independent of the redshift in the range $6 \leq z \leq 12$. Whereas \citet{magg16} find an almost constant rate of PISNe between $z=20$ and $z=6$, \cite{hummel12} and \citet{johnson13} report a steep decrease in the PISN rate after $z\approx 10$ with almost no PISN at $z=6$. The rate of PISNe is to first order proportional to the star formation rate of Pop~III stars and both are strongly suppressed at lower redshift by radiative and chemical feedback. Under the assumptions that no PISNe explode at $z<9$, we derive the optimal diagnostic to identify PISN at $z\geq 9$. We use the same standard set of PISN with 5 different progenitor masses as before, but now consider only the redshifts $z=9$ and $z=12$.


This restriction to PISN at higher redshift increases the required exposure times to $>6$\,ks. The $250\Msun$ PISN is faint between 2 and 4\,yr in the observer frame, which renders the $225\Msun$ PISN at $z=9$ the dominant contribution for the optimal exposure time of $3$\,ks. The steady change of the 200W/277W color with time after explosion makes it easy to detect these PISNe as transient sources. 

\begin{figure}
\centering
\includegraphics[width=0.47\textwidth]{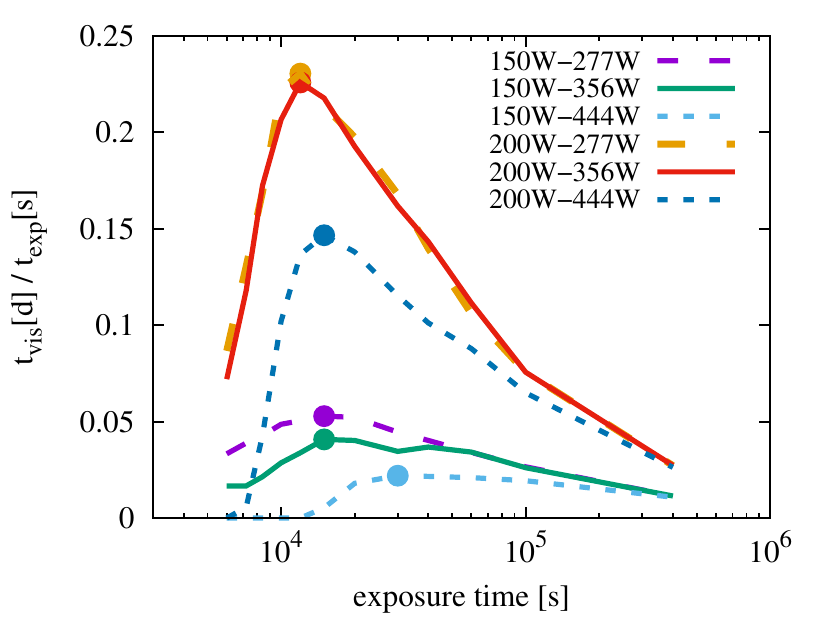}
\caption{Efficiency of given filter combination, measured by the visibility time divided by the exposure time, as a function of the exposure time for PISNe at $z\geq 9$. The F200W-F277W is the most promising diagnostic with an optimal exposure time of 12\,ks.}
\label{fig:texpz9}
\end{figure}

Even for very long exposure times the visible fraction of the lightcurve of $z>9$ PISNe remains below $\sim 60\%$. In Fig. \ref{fig:texpz9} we show the efficiency of these filter diagnostics and derive the optimal exposure time.
The combination of the 200W and 277W filters is the best choice to detect PISNe at $z\geq 9$ with an optimal exposure time of $12$\,ks.

\subsection{Detecting the afterglow}
\label{sec:glow}
\begin{figure}
\centering
\includegraphics[width=0.47\textwidth]{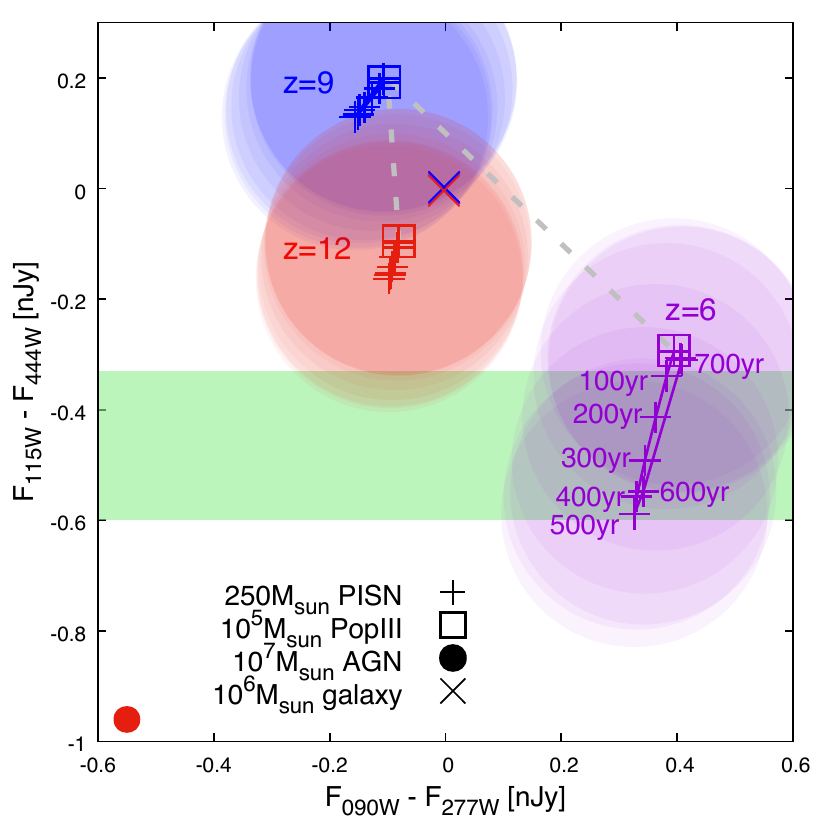}
\caption{Color-color diagnostics to best distinguish PISNe with nebular emission and the echo effect from other high-redshift sources. The three colors illustrate different redshifts and the pluses indicate the time evolution of a $250\Msun$ PISN. \referee{The transparent circles illustrate the observational uncertainty of $\sim 0.23$\,nJy for the $250\Msun$ PISNe at S/N$\geq 100$ in all four filters.} The typical high-redshift galaxy (crosses) is very faint and the AGN (circle) is so bright that only its $z=12$ realisation falls in the relevant color range. In the green range we expect \referee{dominantly} PISNe: for redder colors the high-redshift AGN falls in the same spectral range and for bluer colors, the underlying populations prevents a unique identification (the gray dotted line illustrates the expected smooth color transition with redshift).}
\label{fig:diagEcho}
\end{figure}

We show the most promising 4-filter combinations to detect the PISN afterglow for an ISM density of $n_\mathrm{H}=100\,\mathrm{cm}^{-3}$ in Fig. \ref{fig:diagEcho}.
Evidently, the unique identification of a PISN afterglow is beyond the capabilities of {\it JWST}, because it is too faint to be distinguished from the underlying stellar population. We still wish to outline a possible strategy for detection, with even more powerful telescopes in the future.

The most promising diagnostic is $F_\mathrm{115W}-F_\mathrm{444W}$, because it offers the largest range in which we only expect PISNe (green area). For the $F_\mathrm{115W}-F_\mathrm{277W}$ color, e.g., the $z=12$ AGN is unfavourably close to the PISN signature, which would require an even higher sensitivity to distinguish them.
The $F_\mathrm{090W} - F_\mathrm{F277W}$ color on the horizontal axis cannot be used to discriminate the PISNe from other sources, because the evolution of the squares with redshift has to be seen as a continuous time sequence (gray dotted line).
However, this color could be used as an additional constraint.

For the $F_\mathrm{115W}-F_\mathrm{444W}$ combination, we need a sensitivity of at least $0.2$\,nJy in both filters to discriminate the $250\Msun$ PISN 300-600\,yr after its explosion from the underlying stellar population. This requires an exposure time of $\sim 5$\,yr, far beyond the capabilities of {\it JWST}. \referee{Taking into account the observational uncertainty and the consequently required high signal-to-noise ratio requires even longer exposure times, which makes the identification of a PISN afterglow more challenging.}

The remarkably long visibility time of $\sim 300$\,yr in the source rest frame would be stretched by the cosmic expansion to over $2000$\,yr. However, this yields an efficiency of only 7\,min of visibility per 1\,s invested exposure time and is hence significantly less economical than the detection of the direct PISN emission. Moreover, not every Pop~III hosting halo is expected to exhibit high enough densities to maintain neutral gas and create this afterglow.

This derivation is not intended to guide blind deep field surveys to detect and identify the PISN afterglow. We rather want to illustrate the typical color evolution of such a PISN afterglow, discuss the typical timescales, and highlight which other sources could mimic such a photometric signature. The unique identification of the PISN afterglow is beyond the capabilities of {\it JWST}, but these results are relevant for possible future observations: after the detection of prompt PISN emission, our model predicts the expected afterglow signature, depending on the ISM properties. Hence, the PISN afterglow can be used to probe the ISM at high redshift.

\section{Detection Strategies}

\subsection{Detection Rates}
\label{sec:rates}
With our optimised 2-filter detection strategy (200W-356W, $600$\,s exposure) {\it JWST} can detect PISNe of progenitor masses $225\Msun$ and $250\Msun$ out to redshift $z\sim 7.5$. For a $250\Msun$ PISN we can see $32\%$ of the lightcurve (S/N$>10$ in both filters), which directly translates into the probability of observing the PISN at the right moment. For a $225\Msun$ PISN we can detect $22\%$ of the lightcurve. Assuming a logarithmically flat IMF in the PISN mass range ($140-260\Msun$), the $225\Msun$ and $250\Msun$ PISNe probe together $33\%$ of the PISN mass range. In Table \ref{tab:rate}, we list different PISN rates from the literature.
\begin{table}
 \centering
  \caption{PISN rates per year per {\it JWST} field of view (FoV, $2.2' \times 2.2'$). They roughly agree at $z>7.5$, but differ by almost two orders of magnitude at lower redshift.}
   \label{tab:rate}
 \begin{tabular}{|c|c|c|}
 \hline
  &$z\leq 7.5$ &$7.5 < z \leq 12$\\
    \hline
 \citet{hummel12} & $6 \times 10^{-6}/\mathrm{yr}/\mathrm{FoV}$ & $7 \times 10^{-4}/\mathrm{yr}/\mathrm{FoV}$\\
 \citet{johnson13} & $5 \times 10^{-5}/\mathrm{yr}/\mathrm{FoV}$ & $3 \times 10^{-4}/\mathrm{yr}/\mathrm{FoV}$\\
  \citet{magg16} & $2 \times 10^{-4}/\mathrm{yr}/\mathrm{FoV}$ & $4 \times 10^{-4}/\mathrm{yr}/\mathrm{FoV}$\\
  \hline
  \end{tabular} 
\end{table}
Whereas the different models yield PISN rates, which differ by almost 2 orders of magnitude at $z=6$, they all agree on the expected rates for PISNe from $z\approx 9$. If we assume the optimistic rate for $z<7.5$ by \citet{magg16},  and multiply it with the fraction of the PISN progenitor mass range that is observable and with the probability that we see the PISN lightcurve at the right time, the detection rate for PISNe with {\it JWST} is
\begin{equation}
R_{600\mathrm{s}} = 2 \times 10^{-5}\,\mathrm{yr}^{-1}\,\mathrm{FoV}^{-1}.
\end{equation}
With an optimized survey strategy, {\it JWST} will be able to detect one PISN per year per $50,000$ different exposures of 600\,s each. In the more conservative case, with no PISNe at $z\leq 7.5$, the detection rate will be $(2.7 \pm 0.4) \times 10^{-5}\,\mathrm{yr}^{-1}\,\mathrm{FoV}^{-1}$, but for a longer exposure time of 12\,ks for each individual FoV. This detection rate is in agreement with the mean and variance of the models by \citet{hummel12,johnson13,magg16}. 

The detection rates are proportional to the star formation efficiency of Pop~III stars, and are directly linked to the range and shape of their IMF. Any {\it JWST} constraints will therefore help to reveal the nature of the first stars, and even their non-detection can be used to elucidate the Pop~III IMF \referee{by providing an upper limit on the occurrence of Pop~III PISNe \citep{yoshida04}}.

\subsection{Optimal Detection Strategies}
PISNe are rare events and a total exposure time of at least $3 \times 10^7$\,s, distributed over $50,000$ FoVs, is required to detect at least one event. It is evident that a blind survey, only dedicated to find such PISNe from the first stars, is not feasible. We therefore present other strategies and approaches to detect the first SNe with {\it JWST}.

Archival data helps to select promising targets for follow-up observations. The most unique feature of PISNe is not their extreme luminosity or specific color, but rather their transient nature and shape of their lightcurve \citep{pan12,kozyreva14}. The identification of a PISN candidate and a well-targeted follow-up observation of this source will be one of the most promising paths to identify the first SNe. With the right filter combinations, there should be about one candidate in every 50,000 {\it JWST} FoVs, assuming the rates by \citet{magg16}.

To make this strategy successful, we highlight the optimal filter combinations to identify PISNe. Independent of their redshift distribution, we find the F200W filter to be the optimal choice in the short wavelength channel and the F356W (F277W) filter in the long wavelength channel for PISNe below (above) $z\approx 7.5$. These filters are also required in the most promising 4-filter diagnostics. All observations with $t_\mathrm{exp} \gtrsim 600$\,s, employing these filters, will help in selecting PISN candidates for follow-up.
The latter could be performed with {\it JWST}, or with next-generation ground-based telescopes such as the Giant Magellan Telescope (GMT), or the European Extremely Large Telescope (E-ELT). A main advantage of the first SNe is their cosmologically dilated lightcurve, which makes them observable in follow-ups for over 10 years. Other observational facilities, such as WFIRST, LSST, and Pan-STARRS will complement the search for the first SNe \citep{young08,whalen13,tanaka13,smidt15}.

The overhead of $\sim 280$\,s for the guide star acquisition after pointing to a new target is of the same order as the optimal exposure time to detect PISNe. \referee{Additional indirect overheads can account for several tens of per cent, depending on the exposure time.} For the necessary $\sim 50,000$ individual FoVs, the overhead would sum up to an unnecessary long amount of unused time. We hence rather suggest to include a $600$\,s exposure with the optimal filters F200W and F356W, every time that {\it JWST} points towards a new blind field. \referee{Moreover, parallel observation with NIRCam as secondary instrument will help to increase the sky-coverage with little additional cost.}

A lower threshold than S/N$=10$, yields shorter optimal exposure times and hence a higher probability of identifying PISNe. Especially for searching the archival data for PISN candidates, the desired signal-to-noise ratio can be lowered to identify more interesting candidates for follow-up observations. Also strong gravitational lensing by massive galaxies and galaxy clusters at lower redshift could boost the flux from the first SNe and therefore increase their detection rates
\citep{rydberg13,whalen13}.

The afterglow of a PISN is fainter than the prompt emission and distinguishing it from other sources requires sub-nJy sensitivities, which will not be feasible with {\it JWST}. The more reliable and realistic confirmation of the afterglow is its transient nature on timescales of several hundred years. We can create catalogs of PISNe and their candidates and monitor them over the next decades and even centuries, similar to follow-up studies of historical SN events \citep{somers97,shara12,miszalski16,shara17}. As is the case for GRBs \citep{gehrels09}, this opens a novel and unique technique to study the host environments of the first stars.

\subsection{Caveats}
\label{sec:caveats}
Our model is based on several assumptions and approximations that we want to critically discuss.

The PISN lightcurves employed here are based on the model by \citet{kasen11}, although there are other models, which yield different predictions for the spectral time evolution \citep{scan05,woosley07,pan12,chen15,jerkstrand16,kozyreva17,gilmer17}. A quantitative comparison of the different models is beyond the scope of this paper, and our photometric signatures may be affected by our specific choice \citep{kozyreva17}.
Our optimal survey strategy targets the remnants of Pop~III stars with masses above $200\Msun$. Their existence is not yet confirmed and simulations of primordial star formation tend to predict lower typical masses for the first stars \citep{stacy10,hirano17}. However, the possible non-detection of PISNe over the next decade would not exclude their existence, because Pop~III stars below $200\Msun$ could also yield PISNe, but they might be below our detection limits.

We have included several high-redshift sources that could mimic a PISN, but we cannot exclude all possible objects. Especially tidal disruption events of stars in the strong tidal field near a supermassive black hole have a similar signature and transient nature as a SLSN \citep{ll16,holoien16}, and it is difficult to distinguish them photometrically. \referee{We compare the photometric signature of PISN with our optimized exposure time to other Types of SNe and find that PISN at $z\geq6$ are redder than the considered types of SNe. The unique signature of PISN is their long lightcurves powered by significant amounts of $^{56}$Ni. Especially the most massive PISNe are clearly distinguishable from other SN types by their long rise to maximum luminosity \citep{kozyreva14}. Moreover, the lack of metal lines at early times could provide a unique spectroscopic signature for PISNe \citep{hummel12}.}
In addition, M-dwarfs have a similar red spectrum and their proper motion mimics an intrinsic variability if they disappear at their location in a follow-up survey. A comprehensive analysis of possible local, photometrically similar, sources needs to be carried out in future work.


Although the a-posteriori treatment of the resonant Ly$\alpha$ scattering does not change the photometric signature of the afterglow, it affects its time evolution: the path of Ly$\alpha$ photons is longer due to multiple scatterings, and our predicted time over which we smooth the emission gets also stretched. A self-consistent radiative transfer calculation is better suited to correctly predict the photometric time evolution of the afterglow signature. Also the ISM substructure and SNe that explode off the halo center will alter the reprocessing of the radiation. Although these effects are crucial for a proper characterization of the afterglow, the reprocessed radiation is below the detection limit of {\it JWST}, and only prompt emission will be observable, if at all.

\section{Summary and Conclusions}
We determined the photometric signature of PISNe and derive the optimal filter combination and detection strategy for their detection with {\it JWST}: two wide-band filters F200W and F356W, together with an exposure time of $t_\mathrm{exp}=600$\,s, maximize the visibility time per invested exposure time. The goal is to analyze $>50,000$ different FoVs to select promising PISN candidates \referee{at $z \lesssim 7.5$} that can be further analyzed with follow-up observations for their transient nature. \referee{Consequently, we have to maximize the fraction of the sky covered by JWST with NIRcam. To achieve this, we have to make use of the parallel observation mode of JWST for which our modest requirements are a strong advantage. Every primary exposure with $t_\mathrm{exp}\geq 600$\,s that does not use coordinated parallels and allows for parallel observation with NIRCam is suitable for our purpose. NIRCam can be used as second parallel instrument together with MIRI imaging, NIRISS WFSS, or NIRSpec MOS as primary instrument.}

\referee{Moreover, archival searches will be of great benefit to identify PISN candidates. The results from the guaranteed time observations can be searched for unique signatures of PISNe and then targeted for follow-up observations.}

In parallel to high-redshift gamma-ray bursts \citep{ciardi00,gehrels09}, PISNe will also help us to probe their host environment and the pre-reionization intergalactic medium (IGM) via their afterglow emission. The time evolution of the afterglow is sensitive to the density and radius of the ISM, with the change in magnitude revealing the ionization state of the gas and hence the emissivity of ionizing photons of the underlying stellar population, and the absorption spectroscopy probing properties of the IGM.

The planned launch of {\it JWST} in 2019, together with complementary facilities such as WFIRST or the LSST, will enable the detection of the first SNe, thus ushering in the epoch of direct detections of the first stars. Our novel approach to determine the optimal filter combination and exposure time can also be applied to other surveys to maximize their success. The transient Universe at high redshifts is about to provide an exciting new window into the end of the cosmic dark ages.

\subsection*{Acknowledgements}
\referee{We thank the reviewer for constructive suggestions and careful reading of the manuscript.} We are grateful to Kirk Barrow for providing his SEDs of high-$z$ galaxies, Mattis Magg for sharing his PISN rates, and Dan Kasen for the availability of his PISN SEDs. We appreciate valuable discussions with Caterina Umilta, Alba Vidal Garcia, Simon Glover, Alex Ji, and Eric Pellegrini. TH acknowledges funding under the European Community's Seventh Framework Programme (FP7/2007-2013) via the European Research Council (ERC) Grant `BLACK', under the project number 614199. TH is a JSPS International Research Fellow. VB was supported by NSF grant AST-1413501. AL was supported in part by the Black Hole Initiative at Harvard University, which is funded by a grant from the John Templeton Foundation. Calculations were performed with version 13.04 of {\sc cloudy}, last described by \cite{ferland13}.

\bibliographystyle{mn2e}
\bibliography{PopIII_PISN_JWST}

\bsp
\label{lastpage}

\end{document}